\documentclass[twocolumn,prc,amsmath,amssymb,nofootinbib,preprintnumbers,nolongbibliography]{revtex4-2}

\usepackage[dvipdfmx]{graphicx,color}
\usepackage{amsmath,amssymb}
\usepackage{bm}
\usepackage{textcomp}
\usepackage{ulem}
\usepackage{xcolor}

\newcommand{\nucl}[2]{{}^{#1}\text{#2}}

\begin{document}

\preprint{NITEP 231}

\title{Deformation and core$+n$ decoupling in the spectrum of $\nucl{17}{C}$}

\author{Tadahiro Suhara}
\email{suhara@matsue-ct.ac.jp}
\affiliation{National Institute of Technology (KOSEN), Matsue College, Matsue 690-8518, Japan}

\author{Yasutaka Taniguchi}%
\affiliation{Department of Computer Science, Fukuyama University, Fukuyama 729-0292, Japan}
\affiliation{RIKEN Nishina Center, Wako 351-0198, Japan}

\author{Wataru Horiuchi}%
\affiliation{Department of Physics, Osaka Metropolitan University, Osaka 558-8585, Japan}
\affiliation{Nambu Yoichiro Institute of Theoretical and Experimental Physics (NITEP), Osaka Metropolitan University, Osaka 558-8585, Japan}
\affiliation{RIKEN Nishina Center, Wako 351-0198, Japan}
\affiliation{Department of Physics, Hokkaido University, Sapporo 060-0810, Japan}

\author{Shin Watanabe}%
\affiliation{National Institute of Technology (KOSEN), Gifu College, Motosu 501-0495, Japan}
\affiliation{RIKEN Nishina Center, Wako 351-0198, Japan}

\author{Takenori Furumoto}
\affiliation{College of Education, Yokohama National University, Yokohama 240-8501, Japan}

\date{\today}

\begin{abstract}
\noindent{\bf Background:}
The coexistence of various structures, such as diverse shapes and cluster structures, is a fundamental property of atomic nuclei.
In neutron-rich nuclei, a core$+n$ structure can compete with nuclear deformation due to the small neutron separation energy.
A neutron-rich carbon isotope, $\nucl{17}{C}$, exemplifies the appearance of the deformation and the core+$n$ decoupling in its spectrum, which is desirable for a deeper understanding of the coexistence phenomena in neutron-rich nuclei.

\noindent{\bf Purpose:} 
We aim to describe and understand this coexistence phenomenon in the low-lying levels of $\nucl{17}{C}$ in a unified manner considering explicitly the degrees of freedom of both the quadrupole deformation and the relative motion between a $\nucl{16}{C}$ core and a valence neutron.

\noindent{\bf Method:}
We adopt the generator coordinate method (GCM) with the antisymmetrized molecular dynamics (AMD) to describe various configurations.
We superpose various basis wave functions generated by the energy variation by imposing two types of constraints: one incorporating the degree of the quadrupole deformation and the other taking care of the relative motion between a $\nucl{16}{C}$ core and a valence neutron.

\noindent{\bf Results:}
We find that the experimental energy level is well reproduced by the present method, including both deformed and $\nucl{16}{C}$+$n$ configurations.
The ground $3/2^{+}$ and second excited $5/2^{+}$ states exhibit a triaxially deformed shape, while the main component of the first excited $1/2^{+}$ state is a $\nucl{16}{C}(0^{+}$) core plus an $s$-wave neutron configuration. 
The tail of the valence neutron is significantly improved by including the $\nucl{16}{C}$+$n$ basis functions explicitly.

\noindent{\bf Conclusion:}
The explicit inclusion of both the quadrupole deformation and the relative motion between a core and a valence neutron is essential to describe the coexistence phenomena observed in neutron-rich nuclei in the AMD+GCM framework.
\end{abstract}

\pacs{21.60.Cs, 21.60.Gx, 27.20.+n}

\maketitle

\section{Introduction}\label{introduction}

The coexistence of various structures is a fundamental property of atomic nuclei.
Most ground states exhibit quadrupole deformations, and when low-lying excited states are considered, the coexistence phenomena associated with various shapes appear across a wide region of the nuclear chart~\cite{Wood:1992zza,Kris:2011fad}.
For example, the coexistence of spherical and largely deformed nuclear shapes in Ni isotopes~\cite{Tsunoda:2013hsa}, the prolate-oblate shape in $\nucl{28}{Si}$~\cite{DasGupta:1967nqx,PhysRevLett.46.1559,GSB+81,Sheline1982263,PhysRevC.71.014303,Taniguchi:2009wp} and Se isotopes~\cite{Hinohara:2009kq}, and triple shapes of $\nucl{186}{Pb}$~\cite{Andreyev:2010nat} were suggested.
Other types of coexistence phenomena occur in light nuclei, characterized by various cluster structures.
For example, the coexistence of various molecular and atomic structures in Be isotopes~\cite{Itagaki:1999vm,Descouvemont:2001wek,Ito:2003px,Itagaki:2004up,Ito:2005yy,Ito:2008zza,Kanada-Enyo:2012yif,Otsuka:2022bcf}, various geometrical configurations of $\alpha$ clusters in C isotopes~\cite{Uegaki:1977ptp,Kanada-Enyo:1998onp,Bijker:2000fw,Neff:2003ib,Kanada-Enyo:2006rjf,Freer:2009zz,Suhara:2010ww,Marin-Lambarri:2014zxa,Suhara:2014wua,Baba:2014lsa,Stellin:2015wsq,Baba:2016sbi,Kimura:2016fce,Bijker:2019vql,Vitturi:2019hgm,Casal:2020agw}, and a deformed shape, an $\alpha$+$\nucl{16}{O}$ cluster structure, and a $2\alpha$+$\nucl{12}{C}$ cluster structure in $\nucl{20}{Ne}$ were discussed~\cite{Fujiwara:1980,Taniguchi:2004zz}.
It is known that level schemes and various observables can be explained by incorporating these coexistence phenomena.

In neutron-rich nuclei, a different type of coexistence phenomenon is expected, as both quadrupole deformation and relative motion between a core and a valence neutron can be essential degrees of freedom~\cite{Kanada-Enyo:2004tao,Sagawa:2004ut,Freer2007TheCN}.
The quadrupole deformation appears in many nuclear systems, while in neutron-rich nuclei, one or more valence neutrons can easily decouple from the ``core'' nucleus because of its small neutron separation energy.
This decoupling phenomenon, a core$+Xn$ structure, has often been used as a basic assumption for describing the weakly bound neutron-rich nuclei. 
See, for example, Refs.~\cite{Horiuchi:2006es,Horiuchi:2006ga} for neutron-rich C isotopes.
Over the decades, a deformed core plus valence nucleon system has been studied using macroscopic and semi-microscopic few-body models with effective deformation parameters~\cite{Nunes:1996cyo,Tarutina:2004cpy,Thompson:2004dc,Lay:2012fj}, microscopic inputs~\cite{Lay:2014zwa,Descouvemont:2013qva}, and Nilsson-like orbits~\cite{Punta:2023wup,Hamamoto:2007ht}.
Reliable modeling of deformation and core+n decoupling requires microscopic models, but such studies remain limited.
Actually, in most previous microscopic approaches, quadrupole deformation and core$+Xn$ decoupling have often been treated in different schemes.
However, the simultaneous consideration of these degrees of freedom is important for a unified understanding of the structure of the low-lying spectrum in neutron-rich nuclei.

In this study, we investigate $\nucl{17}{C}$, where experimental observations suggested the possibility of this coexistence phenomenon in its low-lying spectrum~\cite{Suzuki:2008zz,Smalley:2015ngy,Chien:2023iul}.
The ground state (the spin-parity $J^{\pi} = 3/2^{+}$) and the first excited state ($J^{\pi} = 1/2^{+}$) are considered to have a deformed shape and a core$+n$ structure, respectively.
We first discuss the deformation property of the ground state.
When nucleons simply occupy the orbits in ascending order of energy in the spherical shell model, the spin-parity of the ground state of $\nucl{17}{C}$ should be $J^{\pi} = 5/2^{+}$ but is actually observed as $J^{\pi} = 3/2^{+}$~\cite{NNDC}. 
This can be understood by a simple Nilsson model~\cite{Nilsson:1955fn} with prolate deformation, where the valence neutron occupies the $\Omega = 3/2$ orbit originating from the $0d_{5/2}$ shell. 
The Coulomb breakup reaction measurement suggested that the configuration of the ground state of $\nucl{17}{C}$ is a mixture of $\nucl{16}{C}(0^{+}, 2^{+}, 4^{+}) + n$ configurations~\cite{DattaPramanik:2003her}, which is consistent with the ground $3/2^{+}$ state being well deformed.
In contrast, the first excited state of $\nucl{17}{C}$ has $J^{\pi} = 1/2^{+}$.
This cannot be explained by the simple Nilsson model but could be a one-neutron halo structure as the state has very small one-neutron separation energy $S_{n}$, 0.52 MeV~\cite{Suzuki:2008zz}. 
The lifetime measurements found that the $M1$ transition from the first excited $1/2^{+}$ state to the ground $3/2^{+}$ state is strongly hindered~\cite{Suzuki:2008zz,Smalley:2015ngy}.
This means that the $1/2^{+}$ state has a quite different structure from the ground state, supporting the possibility of having a one-neutron halo structure of $\nucl{16}{C}$+$n$.
We remark that theoretical studies using microscopic cluster models predicted that the $1/2^{+}$ state has a one-neutron halo structure~\cite{Timofeyuk:2010zza, Chien:2023iul}.

As we see, the low-lying spectrum of $\nucl{17}{C}$ is an ideal example exhibiting both the nuclear deformation and the core$+n$ decoupling. 
It is desirable to establish a theory that can describe both of them in a unified manner.
For this purpose, we adopt the generator coordinate method (GCM)~\cite{Hill:1952jb,Ring1980NMBP} with the antisymmetrized molecular dynamics (AMD)~\cite{Kanada-Enyo:1994aoa,Kanada-Enyo:1995jpd,Kanada-Enyo:2001yji,Kanada:2003phy,Kanada-Enyo:2012yif,Kimura:2016fce}, abbreviated as AMD+GCM.
AMD is a microscopic structure theory using the antisymmetrized product of Gaussian wave packets, which describe the motion of nucleons. 
In AMD, the nucleon configurations are determined variationally.
By applying an appropriate constraint method, various configurations representing nuclear deformation and clustering can be generated.
For example, one can perform the energy variation with a constraint on quadrupole deformation parameters $\beta$ and $\gamma$. This procedure, the so-called $\beta$-$\gamma$ constraint method, is successful in describing prolate, oblate, and triaxially deformed shapes in a single scheme~\cite{Suhara:2009jb}.
If one takes the distance of the subsystems as a constraint ($d$ constraint method~\cite{Taniguchi:2004zz}), various cluster structures can be generated, including the core$+n$ structures.
Here we superpose those AMD wave functions obtained by both the $\beta$-$\gamma$ and $d$ constraint methods to treat the quadrupole deformation and the core$+n$ decoupling simultaneously.

This paper is organized as follows. 
Section~\ref{formulation} introduces the wave function and Hamiltonian employed in the present study. 
Section~\ref{results} presents calculated energy levels, overlap functions, and $M1$ transition strengths and discusses the coexistence phenomena of the deformation and the core+$n$ decoupling in the spectrum of $\nucl{17}{C}$.
The results for $\nucl{17}{Na}$, which is the mirror nucleus of $\nucl{17}{C}$, are also presented.
Finally, the conclusion is given in Sec.~\ref{conclusion}.

\section{Formulation}\label{formulation}

\subsection{Wave function}

To describe the different features of deformed and core$+n$ decoupled states simultaneously, we superpose several basis wave functions using GCM~\cite{Hill:1952jb,Ring1980NMBP}.
Here, the basis wave functions for GCM are generated by AMD.

In AMD, the basis function of an $A$-nucleon system $|\Phi \rangle$ is expressed by a Slater determinant of single-particle wave functions $|\varphi_{i} \rangle$ as
\begin{equation}
	|\Phi \rangle = \frac{1}{\sqrt{A!}} \det \left\{ |\varphi_{1} \rangle, \cdots ,|\varphi_{A} \rangle \right\}.
\end{equation}
Each single-particle wave function $|\varphi_{i} \rangle$ consists of spatial $|\phi_{i} \rangle$, spin $|\chi_{i} \rangle$, and isospin $|\tau_{i} \rangle$ parts as
\begin{equation}
	|\varphi_{i} \rangle = |\phi_{i} \rangle \otimes |\chi_{i} \rangle \otimes |\tau_{i} \rangle.
\end{equation}
The spatial part $|\phi_{i} \rangle$ is represented by a Gaussian wave packet whose center is $\bm{Z}_{i}$, the spin part $|\chi_{i} \rangle$ is parametrized by $(\xi_{i\uparrow}, \xi_{i\downarrow})$, and isospin part $|\tau_{i} \rangle$ is fixed to be proton or neutron as 
\begin{align}
	\langle \bm{r} | \phi_{i} \rangle &= \left( \frac{2\nu}{\pi} \right)^{\frac{3}{4}} \exp \left[ - \nu \left( \bm{r} - \bm{Z}_{i} \right)^{2} \right], \\
	|\chi_{i} \rangle &= \xi_{i\uparrow} |\uparrow \ \rangle + \xi_{i\downarrow} |\downarrow \ \rangle,\\
	|\tau_{i} \rangle &= |p \rangle \ \text{or} \ |n \rangle.
\end{align}
$\bm{Z}_{i}$, $\xi_{i\uparrow}$, and $\xi_{i\downarrow}$ are complex variational parameters to be determined by the energy variation.
The real and imaginary parts of $\bm{Z}_{i}$ correspond to the position and momentum of the single particle, respectively.
By introducing the imaginary part, we can efficiently describe the $jj$-coupling single-particle wave functions, such as $p_{3/2}$ and $d_{5/2}$~\cite{Itagaki:2005sy,Suhara:2013mja}.
The width parameter $\nu$ is fixed as 0.175~fm$^{-2}$~\cite{Kanada-Enyo:2004tao}.
The calculated root-mean-square radius of $\nucl{16}{C}$ is 2.64~fm, 
which is close to the empirical value $2.70 \pm 0.03$~fm~\cite{Ozawa:2001hb}.

The total wave function of $\nucl{17}{C}$ is expressed by superposing many AMD basis functions based on GCM.
The generator coordinates are set as $\beta$ and $\gamma$, the quadrupole deformation parameters; 
and $d$, the distance between the neutron and the remaining $\nucl{16}{C}$ core.
The quadrupole deformation parameters, $\beta$ and $\gamma$, are defined by
\begin{align}
	\beta \cos \gamma &= \sqrt{\frac{\pi}{5}} 
		\frac{2 \langle z^{2} \rangle - \langle x^{2} \rangle - \langle y^{2} \rangle}{\langle x^{2} \rangle + \langle y^{2} \rangle + \langle z^{2} \rangle}, \\
	\beta \sin \gamma &= \sqrt{\frac{3 \pi}{5}} \frac{\langle x^{2} \rangle - \langle y^{2} \rangle}{\langle x^{2} \rangle + \langle y^{2} \rangle + \langle z^{2} \rangle}.
\end{align}
The distance $d$ is defined as the center of mass position of two quasi clusters, C$_1$ and C$_2$:
\begin{equation}
	d = \lvert \bm{R}_{1} - \bm{R}_{2} \rvert
\end{equation}
with the center of mass position of the quasi clusters C$_{n}$ with the mass number $A_n$:
\begin{equation}
	\bm{R}_{n} = \frac{1}{A_{n}} \sum_{i \in \text{C}_{n}} \text{Re} \bm{Z}_{i}.
\end{equation}
To describe the core$+n$ decoupling, we assign C$_{1}$ and C$_{2}$ to a $\nucl{16}{C}$ core and the remaining neutron, respectively. 
Note that the $\nucl{16}{C}$ core can be excited to a mixture of $J^{\pi} = 0^{+}$, $2^{+}$, and $4^{+}$ states in this method.

In the present computation, we take energy variations for each basis function with
the $\beta$-$\gamma$~\cite{Suhara:2009jb} and $d$~\cite{Taniguchi:2004zz} constraint methods.
For the quadrupole deformation parameters, we take $\beta \le 0.5$ and $ 0^{\circ} \le \gamma \le 60^{\circ}$ on the $\beta$-$\gamma$ plane at 66 mesh points of the triangle lattice.
For the distance $d$, we perform variational calculations for 1--15~fm with an interval of 1~fm and obtain 15 basis wave functions.
We confirmed the convergence of the level ordering and the excitation energies by doubling the mesh width, though there is a minor energy shift of approximately 0.3~MeV overall.

Finally, the total wave function with the spin-parity $J^\pi$ is given by
\begin{equation}
	|\Psi ^{J^{\pi}}_M \rangle = 
	\sum_{iK} f_{\beta_{i} \gamma_{i} K} P^{J \pi}_{MK} |\Phi_{\beta_{i} \gamma_{i}} \rangle
	+ \sum_{iK} f_{d_{i} K} P^{J \pi}_{MK} |\Phi_{d_{i}} \rangle, 
\end{equation}
where $|\Phi_{\beta_{i} \gamma_{i}} \rangle$ and $|\Phi_{d_{i}} \rangle$ are the basis functions obtained by $\beta$-$\gamma$ and $d$ constraint methods, respectively, and $P^{J \pi}_{MK}$ is the parity and angular momentum projection operator.
The coefficients $f_{\beta_{i} \gamma_{i} K}$ and $f_{d_{i} K}$ are determined by diagonalizing the norm and Hamiltonian matrices.
In this paper, we examine three different model spaces to clarify the effects of the quadrupole deformation and the core$+n$ decoupling: the $\beta$-$\gamma$ model space including only the $\beta$-$\gamma$ bases, the $d$ model space including only the $d$ bases, and the full model space including both the $\beta$-$\gamma$ and $d$ bases.
Since our calculations adopt the bound-state approximation, unphysical pseudostates may appear. However, these pseudostates are located above 10~MeV, which does not coexist in the physical low-lying excited states discussed in this paper.

\subsection{Overlap function and spectroscopic factor}

To extract the $\nucl{16}{C}+n$ components from the $\nucl{17}{C}$ wave function, we calculate the overlap function between the $\nucl{17}{C}$ state and the $\nucl{16}{C}+n$ wave functions.
The overlap function is defined as
\begin{equation}
	\mathcal{F}^{J^\pi}_{c} (r) = \frac{1}{r} \big\langle 
		\big[ \Psi^{I^{+}}_{^{16}\text{C}} [Y_{l}(\hat{r}) \chi_{1/2}]^{j} \big]^{J}_M
		\big| \Psi ^{J{\pi}}_M \big\rangle,
\end{equation}
where $r$ is the radius measured from the center-of-mass of the $\nucl{16}{C}$ core to the position of the valence neutron, $\Psi^{I^{+}}_{^{16}\text{C}}$ is a $\nucl{16}{C}$ wave function of the spin parity $I^{+}$ state, $\chi_{1/2}$ is a neutron spinor, index $l$ and $j$ respectively stand for the orbital and total angular momentum of the neutron relative to the $\nucl{16}{C}$ core, and index $c$ stands for $(I^{+}, l, j)$.
$(l,j)$ can be expressed in another way as the single particle orbits of the neutron, e.g., $s_{1/2}$, $d_{3/2}$, and $d_{5/2}$.
For the index $c$, we use the notation with the direct product of $I^{+}$ and the single particle orbit of the neutron, for example, $0^{+} \otimes s_{1/2}$.
The wave function of $\nucl{16}{C}$ is approximated by a single $I^{+}$ projected Slater determinant.
The squared overlap between this representative single Slater determinant and the original $\nucl{16}{C}$ wave function is more than 91\% for $0^{+}$, $2^{+}$, and $4^{+}$ states.
Using the overlap function, the spectroscopic factor is defined as
\begin{equation}
	S^{J^\pi}_{c} = \int_{0}^{\infty} dr \big| \mathcal{F}^{J^\pi}_{c} (r) \big|^{2}.
\end{equation}
Note that due to the approximation of $\nucl{16}{C}$, the overlap function values may change at most by about 9\% but it does not affect the discussions and conclusions.

\subsection{Hamiltonian}

We take the same functional form of Hamiltonian as used in Refs.~\cite{Suhara:2009jb, Suhara:2010ww,Suhara:2012zr,Suhara:2013mja,Furumoto:2013eaa,Furumoto:2017xga,Furumoto:2021jxd,Furumoto:2023bgq}. 
The Hamiltonian consists of the kinetic term, the two-body effective nuclear force, and the Coulomb force.
We adopt the Volkov No.~2 interaction~\cite{Volkov:1965zz} for the central nuclear force.
The spin-orbit force of the G3RS interaction~\cite{Yamaguchi:1979hf} is implemented.
This Hamiltonian has the advantage of reproducing the deformation property of the $\nucl{16}{C}$ core.
We set the Majorana exchange parameter $M = 0.579$, the Bartlett and Heisenberg exchange parameters $B = H = 0$ in the central force, and $V_{\text{LS}} = 800$~MeV in the spin-orbit force. 
With these parameters, the energy levels of $J^\pi=0^{+}$, $2^{+}$, and $4^{+}$ states of $\nucl{16}{C}$ are reproduced by the GCM calculation with the basis functions obtained by the $\beta$-$\gamma$ constraint method.
The calculated (experimental~\cite{NNDC}) values of the energies for the $0^{+}$, $2^{+}$, and $4^{+}$ states are $-110.53$ ($-110.75$), $-108.93$ ($-108.98$), and $-106.72$ ($-106.61$)~MeV, respectively.

\section{Results}\label{results}

\subsection{Energy levels}

\begin{figure}[tb]
	\centering
	\includegraphics[width=\linewidth]{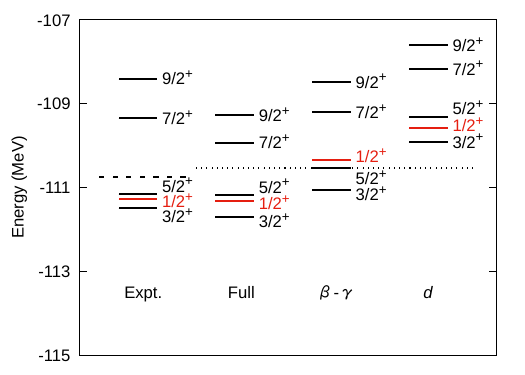}
	\caption{
	Low-lying energy levels of $\nucl{17}{C}$. 
	``Expt.'' indicates the experimental results~\cite{NNDC}.
	Theoretical results are denoted by ``Full'', ``$\beta$-$\gamma$'', and ``$d$''.
    The results of ``$\beta$-$\gamma$'' and ``$d$'' are obtained only with the basis wave functions of the $\beta$-$\gamma$ constraint and the $d$ constraint methods, respectively.
	The result of ``Full'' is obtained with all of them. 
    The dashed and dotted lines indicate the experimental and theoretical $\nucl{16}{C}$+$n$ threshold energies, respectively.
	}
	\label{fig:levels_17C}
\end{figure}

Figure~\ref{fig:levels_17C} displays the low-lying energy levels of $\nucl{17}{C}$.
We show the three types of theoretical results.
The experimental total energies are also shown for comparison as indicated by ``Expt.''.
Experimentally, the spin-parity of the ground state is $J^{\pi} = 3/2^{+}$, and the excited $1/2^{+}$ and $5/2^{+}$ states are observed to be almost degenerate with the ground state.
Our calculation indicated by ``Full'' well reproduces the ground-state energy and the level ordering.
In contrast, calculations indicated by ``$\beta$-$\gamma$'' and ``$d$'' fail to reproduce the level scheme of the low-lying states.
The calculation with the $\beta$-$\gamma$ model space yields the incorrect order of the energy levels.
The order of the $1/2^{+}$ and $5/2^{+}$ states is reversed compared to the experimental data.
The level ordering is well reproduced in the calculations with the $d$ model space, but the total energies are so high that even the lowest-energy state, the $3/2^{+}$ state, becomes unbound.
The reproduction of the level ordering with the $d$ model space supports the validity of macroscopic and semi-microscopic few-body models~\cite{Nunes:1996cyo,Tarutina:2004cpy,Thompson:2004dc,Lay:2012fj,Lay:2014zwa,Descouvemont:2013qva, Punta:2023wup,Hamamoto:2007ht}.
However, in our microscopic model, the binding energy and the level scheme are reproduced when we consider both the $\beta$-$\gamma$ and $d$ model spaces. 
This result indicates that it is necessary to take into account both the deformation and core$+n$ decoupling for a unified understanding of the low-lying states of $\nucl{17}{C}$.

\begin{figure}[tb]
	\centering
	\includegraphics[width=\linewidth]{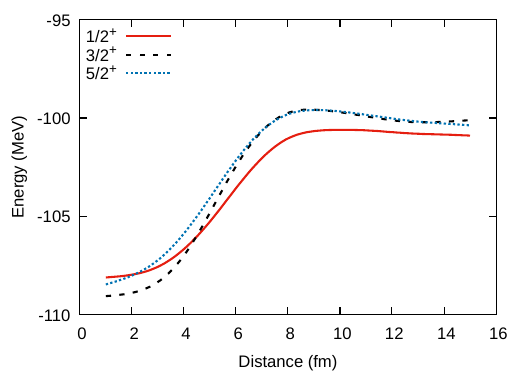}
	\caption{
	Energy curves of the basis wave function for the $1/2^{+}$, $3/2^{+}$, and $5/2^{+}$ states as a function of the distance between the $\nucl{16}{C}$ core and the valence neutron obtained by the $d$ constraint method.
	}
	\label{fig:17C_energy_surface}
\end{figure}

To clarify the mechanism of the interchange of the $5/2^{+}$ and $1/2^{+}$ states by the $\nucl{16}{C}+n$ structure, we show the energy curves for the $1/2^{+}$, $3/2^{+}$, and $5/2^{+}$ states of the basis wave functions obtained by the $d$ constraint method in Fig.~\ref{fig:17C_energy_surface}.
In small $d$, all the $J^{\pi}$ states give the lowest energy and their energy increases as the distance becomes large.
Depending on the distance, the order of the $1/2^{+}$, $3/2^{+}$, and $5/2^{+}$ energies changes.
In $d \lesssim 2$~fm, the $3/2^{+}$ state has the lowest energy and the $5/2^{+}$ and $1/2^{+}$ states appear in this order.
This is the same ordering as the energy levels calculated with the $\beta$-$\gamma$ model space shown in Fig.~\ref{fig:levels_17C}, which is inconsistent with the experimental ordering.
However, the slope of the $1/2^{+}$ energy curve is more gradual than the others, and the $1/2^{+}$ state becomes the lowest for $d \gtrsim 5$~fm.
As shown later in the analysis using overlap functions, by mixing the $d$ constraint basis wave functions, the tail of the wave function of $1/2^{+}$ state is extended to the $d \gtrsim 5$~fm region, leading to the lowering of the $1/2^{+}$ state in the energy levels calculated in the full model space (Fig.~\ref{fig:levels_17C}).
The energy curves are peaked at $d \approx 8$~fm and decrease for $d \gtrsim 8$~fm.
This asymptotic behavior at large $d$ can be explained as follows.
For large $d$, because the interaction between the $\nucl{16}{C}$ core and the neutron disappears, the $\nucl{16}{C}$ core is dominated by the ground $0^{+}$ state. 
In this case, the remaining neutron orbit must coincide with $J^{\pi}$ of $\nucl{17}{C}$, that is, the $s_{1/2}$, $d_{3/2}$, and $d_{5/2}$ orbits correspond to the $J^{\pi} = 1/2^{+}$, $3/2^{+}$, and $5/2^{+}$ states of $\nucl{17}{C}$, respectively. 
Because the $s_{1/2}$ single particle state has no centrifugal potential, the energy of the $1/2^{+}$ state of $\nucl{17}{C}$ becomes smaller than the $3/2^{+}$ and $5/2^{+}$ states. 
We also note that the energies of the $3/2^{+}$ and $5/2^{+}$ states are almost the same at large $d$, as their valence neutron orbits have the same orbital angular momentum, $l=2$ (the $d$ orbit), and the spin-orbit force between the $\nucl{16}{C}$ core and the valence neutron is negligible.
When the $^{16}$C core and the neutron come closer at the intermediate distance $d \approx 8$~fm where the peaks are located, the energies increase because the core$+n$ interaction mixes the $s$ and $d$ orbits, which enables the $\nucl{16}{C}$ core to excite to the $2^{+}$ state.

\subsection{Spectroscopic factors and overlap functions for $\nucl{16}{C}+n$ configurations}

\begin{table}[tb]
	\caption{
	Spectroscopic factors $S^{J^\pi}_{c}$ for the $1/2^{+}$, $3/2^{+}$, and $5/2^{+}$ states in the full and the $\beta$-$\gamma$ model spaces.
    All the configurations for the $s_{1/2}$, $d_{3/2}$, and $d_{5/2}$ valence neutron orbits are listed.
	The index $c$ is a direct product of the spin parity $I^{+}$ of the $\nucl{16}{C}$ core and the single particle orbit of the valence neutron.
	}
	\label{table:17C_S-factor_full}
	\begin{tabular}{c|ccc} \hline \hline
		$J^{\pi}$ & $c$ & $S^{J^\pi}_{c}$(Full) & $S^{J^\pi}_{c}$($\beta$-$\gamma$) \\ \hline
		$1/2^{+}$ & $0^{+} \otimes s_{1/2}$ & 0.521 & 0.481 \\
			 & $2^{+} \otimes d_{3/2}$ & 0.064 & 0.079 \\
			 & $2^{+} \otimes d_{5/2}$ & 0.391 & 0.413 \\
		$3/2^{+}$ & $2^{+} \otimes s_{1/2}$ & 0.139 & 0.121 \\
			 & $0^{+} \otimes d_{3/2}$ & 0.020 & 0.023 \\
			 & $2^{+} \otimes d_{3/2}$ & 0.023 & 0.042 \\
			 & $2^{+} \otimes d_{5/2}$ & 1.252 & 1.300 \\
			 & $4^{+} \otimes d_{5/2}$ & 0.175 & 0.179 \\
		$5/2^{+}$ & $2^{+} \otimes s_{1/2}$ & 0.035 & 0.058 \\
			 & $2^{+} \otimes d_{3/2}$ & 0.071 & 0.071 \\
			 & $4^{+} \otimes d_{3/2}$ & 0.014 & 0.022 \\
              & $0^{+} \otimes d_{5/2}$ & 0.623 & 0.660 \\
			 & $2^{+} \otimes d_{5/2}$ & 0.116 & 0.087 \\
			 & $4^{+} \otimes d_{5/2}$ & 0.643 & 0.654 \\
		\hline \hline
	\end{tabular}
\end{table}

Here we discuss the role of the $\nucl{16}{C}+n$ configurations in the low-lying states of $\nucl{17}{C}$.
Table~\ref{table:17C_S-factor_full} lists the spectroscopic factors $S^{J^\pi}_{c}$ of the $J^{\pi} = 1/2^{+}$, $3/2^{+}$, and $5/2^{+}$ states with the $s_{1/2}$, $d_{3/2}$, and $d_{5/2}$ valence neutron orbits in the full and the $\beta$-$\gamma$ model spaces.
Only the $1/2^{+}$ state contains a sizable $s_{1/2}$ component as indicated by $0^{+} \otimes s_{1/2}$.
The $S^{1/2^{+}}_{0^{+} \otimes s_{1/2}}$ value in the full model space is larger than that in the $\beta$-$\gamma$ model space, which shows that the basis wave functions obtained by the $d$ constraint method enhance the $0^{+} \otimes s_{1/2}$ component.
The $3/2^{+}$ and $5/2^{+}$ states are dominated by the configurations consisting of the $d_{5/2}$ valence neutron orbits, i.e., the $2^{+} \otimes d_{5/2}$ configuration for the $3/2^{+}$ state and the $0^{+} \otimes d_{5/2}$ and $4^{+} \otimes d_{5/2}$ configurations for the $5/2^{+}$ state.
It is consistent with the interpretation of the Nilsson model with prolate deformation, in which the valence neutron occupies the $\Omega = 3/2$ orbit originating from the $0d_{5/2}$ shell.
The dominance of the $2^{+} \otimes d_{5/2}$ configuration in the ground $3/2^{+}$ state is consistent with the $\nucl{16}{C}$+$n$ cluster model result~\cite{Chien:2023iul} and the Coulomb breakup reaction experiment~\cite{DattaPramanik:2003her}.
The $d_{3/2}$ components are found to be small, $S^{J^{\pi}}_{c}<0.10$, for all states.

\begin{figure}[tb]
	\centering
	\includegraphics[width=\linewidth]{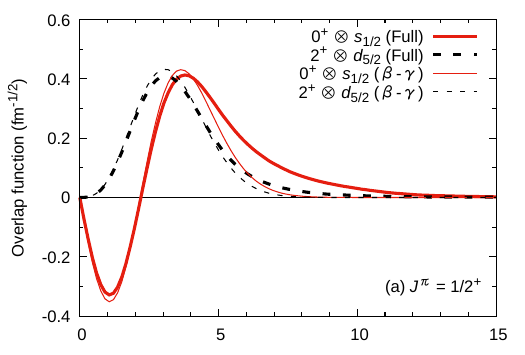}
	\includegraphics[width=\linewidth]{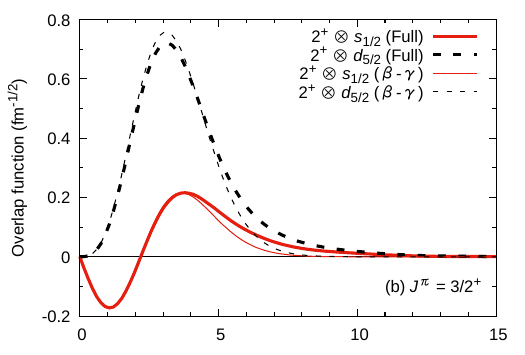}
	\includegraphics[width=\linewidth]{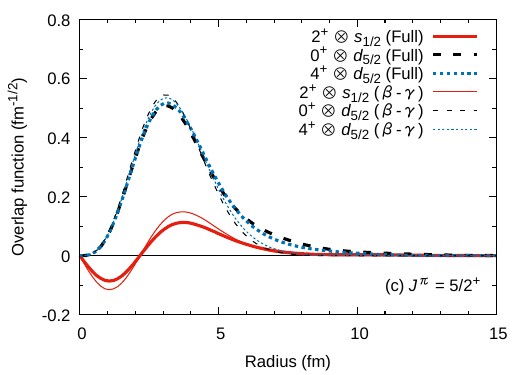}
	\caption{
	Overlap functions of (a) $1/2^{+}$, (b) $3/2^{+}$, and (c) $5/2^{+}$ states of $\nucl{17}{C}$ with the $\nucl{16}{C}+n$ states.
	In the legend, $0^{+}$, $2^{+}$, and $4^{+}$ indicate the yrast states of $\nucl{16}{C}$.
	$s_{1/2}$ and $d_{5/2}$ are the single particle orbits of the valence neutron in the $\nucl{16}{C}+n$ system.
	The labels ``Full'' and ``$\beta$-$\gamma$'' mean the same as Fig.~\ref{fig:levels_17C}.
	}
	\label{fig:17C_overlap_function}
\end{figure}

Figure~\ref{fig:17C_overlap_function} draws the overlap functions of (a) $1/2^{+}$, (b) $3/2^{+}$, and (c) $5/2^{+}$ states of $\nucl{17}{C}$ with the $^{16}\text{C}+n$ states in the full and the $\beta$-$\gamma$ model spaces.
Only the configurations with $S^{J^{\pi}}_{c}>0.20$ among those listed in Table~\ref{table:17C_S-factor_full} are plotted.
Let us discuss the $1/2^{+}$ state shown in Fig.~\ref{fig:17C_overlap_function}(a).
For the $0^{+} \otimes s_{1/2}$ configuration, the two overlap functions are almost identical below $r \approx 4$~fm, while we see that the tail behavior of the overlap function is significantly improved: the overlap function calculated with the full model space has a longer tail compared to that with the $\beta$-$\gamma$ model space.
The basis wave functions with the $d$ constraint method improve the $\nucl{17}{C}$ wave function at the tail region.
As discussed in Fig.~\ref{fig:17C_energy_surface}, the energy curve of the $1/2^{+}$ state becomes the lowest for $d \gtrsim 5$~fm, which is the tail region, resulting in the large mixing of the $\nucl{16}{C}(0^{+})+n$ configuration.
We note, however, that the spectroscopic factor $S^{1/2^{+}}_{0^{+} \otimes s_{1/2}}$ changes only slightly with this improvement even though the difference in the tail is quite large.
Compared to the $0^{+} \otimes s_{1/2}$ configuration, the change of the overlap function in the $2^{+} \otimes d_{5/2}$ configuration is smaller.
This is due to the difference in the relative motion of the valence neutron orbit: the $s$ orbit is free from the centrifugal potential, whereas the $d$ orbit has the centrifugal barrier of about 2 MeV at $r \approx 8$~fm.

The behavior of the wave functions for the $3/2^{+}$ and $5/2^{+}$ states is different from that of the $1/2^{+}$ state.
As shown in Fig.~\ref{fig:17C_overlap_function}(b) for the $3/2^{+}$ state, the tail behavior of the $2^{+} \otimes s_{1/2}$ configuration is improved similarly to the case of the $0^{+} \otimes s_{1/2}$ configuration in the $1/2^{+}$ state, but its impact is minor as the spectroscopic factor $S^{3/2^+}_{2^+\otimes s_{1/2}}$ is small (Table~\ref{table:17C_S-factor_full}).
Actually, the energy gain for the $3/2^{+}$ state from the $2^{+} \otimes s_{1/2}$ configuration is smaller than that for the $1/2^{+}$ state from the $0^{+} \otimes s_{1/2}$ configuration.
The tail of the dominant $2^{+} \otimes d_{5/2}$ configuration shows a smaller change than the $0^{+} \otimes s_{1/2}$ configuration in the $1/2^{+}$ state due to the centrifugal potential.
As shown in Fig.~\ref{fig:17C_overlap_function}(c) for the $5/2^{+}$ state, the basis wave functions with the $d$ constraint method do not have extended tails in the dominant $0^{+} \otimes d_{5/2}$ and $4^{+} \otimes d_{5/2}$ configurations significantly, and the energy gain is also small.
This selective energy gain of the $1/2^{+}$ state, which is due to the improvement of the tail of the $0^{+} \otimes s_{1/2}$ configuration, is essential to reproduce the experimental level ordering of $3/2^{+}$, $1/2^{+}$, and $5/2^{+}$ as shown in Fig.~\ref{fig:levels_17C}.
Note that, in the neutron removal reactions, these tail changes for the $d_{5/2}$ component could be crucial even though they are not important for the spectroscopic factor or the energy gain~\cite{Minomo:2011bb,Watanabe:2014zea}. 

\begin{figure}[tb]
	\centering
	\includegraphics[width=\linewidth]{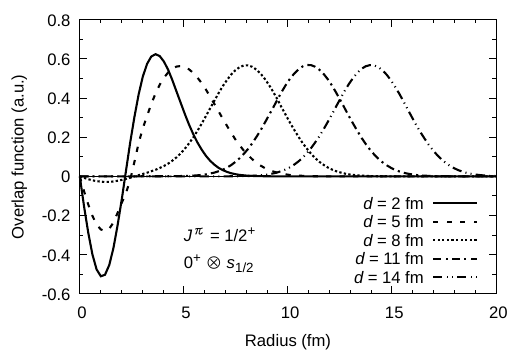}
	\caption{
	Overlap functions for the single $1/2^{+}$ projected basis wave functions obtained by the $d$ constraint method with the $\nucl{16}{C}+n$ wave function. 
	}
	\label{fig:17C_base_overlap_function}
\end{figure}

To explain why the tail is extended in the full model space compared to that in the $\beta$-$\gamma$ model space, Fig.~\ref{fig:17C_base_overlap_function} draws the overlap functions for the single $1/2^{+}$ projected basis wave functions obtained by the $d$ constraint method with the $\nucl{16}{C}+n$ wave function.
The overlap functions are normalized so that the spectroscopic factor $S^{1/2^+}_{0^+\otimes s_{1/2}}$ is unity for the sake of comparison.
For the basis wave functions with $d \gtrsim 5$~fm, the outer part of the overlap function has a single Gaussian type with a peak near $d$.
This means that superposing these basis wave functions is equivalent to a local Gaussian expansion of the tail part of the wave function, making it possible to describe a long tail and improve the wave function by mixing the basis wave functions with large $d$.

The tail behaviors of these different $J^{\pi}$ states represent the deformation and core$+n$ decoupling.
The short neutron tails of the $3/2^{+}$ and $5/2^{+}$ states indicate a strong coupling between a $\nucl{16}{C}$ core and a valence neutron, which can constitute a deformed nuclear shape.
To analyze the shape, we calculate the squared overlap $O$ between the total wave function and the deformed basis wave function,
\begin{equation}
    O = \lvert \langle \Psi ^{J^{\pi}}_M | \Phi_{\beta \gamma} \rangle \rvert^{2}.
\end{equation}
The maximum squared overlaps are observed at the same deformed basis, $(\beta, \gamma) = (0.26, 19^{\circ})$, for both the $3/2^{+}$ and $5/2^{+}$ states, with values of 87.3\% and 84.3\%, respectively, indicating the $3/2^{+}$ and $5/2^{+}$ states have a triaxially deformed nuclear shape.

\subsection{Magnetic dipole ($M1$) transition} \label{subsection:M1_tansition}

Let us discuss the $M1$ transition strength from the $1/2^{+}$ state to the $3/2^{+}$ state.
The strongly hindered experimental value $B(M1; 1/2^{+} \rightarrow 3/2^{+})_{\text{Expt.}} = 1.04^{+0.03}_{-0.12} \times 10^{-2} \mu_{N}^{2}$ was measured, suggesting a one-neutron halo structure of the $1/2^{+}$ state.
The calculation with the $\beta$-$\gamma$ model space yields $B(M1; 1/2^{+} \rightarrow 3/2^{+})_{\text{$\beta$-$\gamma$}} = 5.06 \times 10^{-1} \mu_{N}^{2}$, and that with the full model space yields smaller value $B(M1; 1/2^{+} \rightarrow 3/2^{+})_{\text{Full}} = 3.72 \times 10^{-1} \mu_{N}^{2}$, which approaches the experimental value.
The extension of the $\nucl{16}{C}+n$ tail achieved by the basis wave functions with the $d$ constraint method contributes to this improvement.
However, the theoretical $M1$ transition strength still overestimates the experimental value, which may be attributed to the overestimation of the neutron separation energy: the experimental value of the neutron separation energy $S_{n}$ is 0.52 MeV, but our calculation gives a value of 0.80 MeV.
The overestimation of the $S_{n}$ shrinks the neutron orbits in the core$+n$ configuration, which is more significant for the $s_{1/2}$ orbit as the valence neutron feels no centrifugal potential.
Therefore, with the correct $S_{n}$, we expect significant changes in the structure of the $1/2^{+}$ state, which is dominated by the $0^+\otimes s_{1/2}$ configuration.
With more extended $\nucl{16}{C}+n$ configurations, the spatial overlap between the $3/2^{+}$ and $1/2^{+}$ states would decrease, thereby reducing the $M1$ transition strength. 
For more quantitative studies, it is necessary to use more appropriate effective nuclear forces that reproduce the neutron separation energy.
For example, if we use the same type of interaction employed in this paper, there is room for adjusting the Bartlett and Heisenberg parameters, which are currently set to 0. 
Another possibility is to introduce three-body interactions as was used in Refs.~\cite{Tohsaki:1994zz,Itagaki:2016bxb}.

The calculated $M1$ transitions from the $5/2^{+}$ state to the $3/2^{+}$ state are $B(M1; 5/2^{+} \rightarrow 3/2^{+})_{\text{$\beta$-$\gamma$}} = 2.95 \times 10^{-2} \mu_{N}^{2}$ and $B(M1; 5/2^{+} \rightarrow 3/2^{+})_{\text{Full}} = 9.11 \times 10^{-3} \mu_{N}^{2}$, which are much smaller than the experimental value $B(M1; 5/2^{+} \rightarrow 3/2^{+})_{\text{Expt.}} = 7.12^{+1.27}_{-0.96} \times 10^{-2} \mu_{N}^{2}$~\cite{Smalley:2015ngy}.
This underestimation may be caused by the insufficient neutron $d_{3/2}$ component in the GCM basis wave functions. 
The $\nucl{16}{C}$ core in $\nucl{17}{C}$ can mainly be $0^{+}$, $2^{+}$, and $4^{+}$ states, between which $M1$ transitions are forbidden.
Therefore, ignoring the recoil effect of the core, the $M1$ transition is governed only by the spin flip of the valence neutron.
In this study, we performed the energy variation to calculate the GCM basis wave functions, and therefore the $d_{5/2}$ orbit is preferred to gain the spin-orbit force. 
Actually, the valence neutron orbit is dominated by the $d_{5/2}$ orbit for the $3/2^{+}$ and $5/2^{+}$ states (Table~\ref{table:17C_S-factor_full}). 
Mixing of the $d_{3/2}$ component may enhance the $M1$ transition strength from the $5/2^{+}$ state to the $3/2^{+}$ state due to the strong $M1$ transition between the $d_{3/2}$ and $d_{5/2}$ orbits.
A wider model space including the $d_{3/2}$ component is necessary to reproduce the observed $M1$ transition strength.
To enhance the $d_{3/2}$ component, it is helpful to explicitly incorporate GCM bases in which the spin of the valence neutron is flipped in the $d$ bases.

\subsection{Mirror nucleus $\nucl{17}{Na}$}

\begin{figure}[tb]
	\centering
	\includegraphics[width=\linewidth]{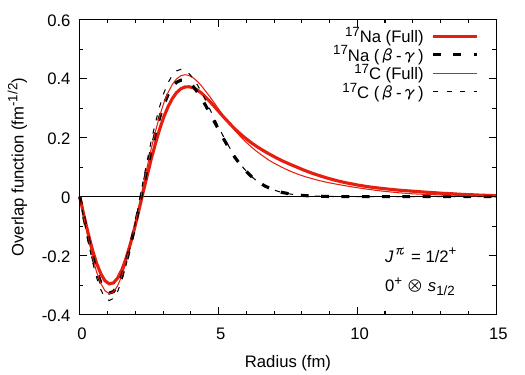}
	\caption{
	Comparison between overlap functions for the $J^{\pi} = 1/2^{+}$ state of $\nucl{17}{C}$ and $\nucl{17}{Na}$ with $c = 0^{+} \otimes s_{1/2}$.
	The labels ``Full'' and ``$\beta$-$\gamma$'' mean the same as Fig.~\ref{fig:levels_17C}.
	}
	\label{fig:17C17Na_overlap_function_0.5+_compare}
\end{figure}

The same calculations are also performed for the mirror nucleus $\nucl{17}{Na}$.
The level scheme of the low-lying states of $\nucl{17}{Na}$ is similar to that of $\nucl{17}{C}$.
For the $5/2^{+}$ state, the excitation energies are the same for $\nucl{17}{Na}$ and $\nucl{17}{C}$.
In contrast, the excitation energy of the $1/2^{+}$ state of $\nucl{17}{Na}$ is 0.08 MeV lower than that of $\nucl{17}{C}$, which is nothing but the Thomas-Ehrman shift~\cite{Thomas:1952,Ehrman:1951}.
Figure~\ref{fig:17C17Na_overlap_function_0.5+_compare} draws the overlap functions for the $J^{\pi} = 1/2^{+}$ states of $\nucl{17}{C}$ and $\nucl{17}{Na}$ with $c = 0^{+} \otimes s_{1/2}$, which is the dominant component of these states.
We see that, in the $\beta$-$\gamma$ model space, the overlap functions for $\nucl{17}{C}$ and $\nucl{17}{Na}$ are almost identical for $r \gtrsim 5$~fm.
In contrast, in the full model space, the tail of $\nucl{17}{Na}$ is more extended than that of $\nucl{17}{C}$.
Since the valence proton in the $s_{1/2}$ orbit has no centrifugal potential, the wave function can be pushed further outward due to the Coulomb repulsion from the $\nucl{16}{Ne}$ core.
The basis wave functions obtained by the $d$ constraint method contribute more to the dilute region around the surface, which is essential to describe the halo and clustering features of atomic nuclei.

\section{Conclusion}\label{conclusion}

In the AMD+GCM framework, we have succeeded in describing the coexistence phenomenon of deformation and core+$n$ decoupling in the spectrum of $\nucl{17}{C}$ by explicitly treating the quadrupole deformation and the relative motion between a core and a valence nucleon simultaneously.
This has been achieved by combining two constraint methods: the $\beta$-$\gamma$ constraint method and the $d$ constraint method.
The ground $J^{\pi} = 3/2^{+}$ and second excited $5/2^{+}$ states of $\nucl{17}{C}$ are mainly described by a triaxially deformed wave function, which is obtained by the $\beta$-$\gamma$ constraint method.
The tail of the valence neutron in the first excited $1/2^{+}$ state of $\nucl{17}{C}$ is significantly improved by including explicitly the $\nucl{16}{C}$+$n$ basis functions, which are obtained by the $d$ constraint method.
It is interesting to apply this method to other neutron-rich nuclei, e.g., $\nucl{19}{C}$ and $\nucl{21}{C}$, in which the deformation and core$+n$ decoupling may coexist in their spectra.

\section*{Acknowledgments} 

This work was supported by the Research Network Support Program, National Institute of Technology (KOSEN) and Japan Society for the Promotion of Science (JSPS) KAKENHI Grant Numbers JP18K03635, JP20K03944, JP21K03543, JP22K03610, JP22K14043, JP22H01214, and JP23K22485.

%

\end{document}